\documentclass[aps,twocolumn,showpacs]{revtex4}
\input epsf

\newcommand{\be}{\begin{equation}}
\newcommand{\ee}{\end{equation}}
\newcommand{\tphi}{\tilde{\phi}}
\newcommand{\beq}{\begin{equation}}
\newcommand{\eeq}{\end{equation}}
\newcommand{\bea}{\begin{eqnarray}}
\newcommand{\eea}{\end{eqnarray}}
\newcommand{\beal}{\setcounter{letter}{1} \begin{eqnarray}}
\newcommand{\eeal}{\addtocounter{equation}{1} \end{eqnarray}}

\begin{document}


\title{\bf Dimension-Dependence of the Critical Exponent in
Spherically Symmetric Gravitational Collapse}

\author{J. Bland${}^\sharp$ and B. Preston${}^\sharp$, M. Becker${}^\sharp$,
 G. Kunstatter$^\sharp$,
V. Husain${}^\flat$}

\affiliation{$\flat$Dept. of Mathematics and Statistics\\
University of New Brunswick,
Fredericton, N.B. Canada E3B 1S5.}

\affiliation{
$\sharp$Dept. of Physics and Winnipeg Institute of
Theoretical Physics\\
University of Winnipeg,
Winnipeg, Manitoba Canada R3B 2E9.}

\date{\today}

\begin{abstract}
We study the critical behaviour of spherically symmetric scalar field collapse to black holes in spacetime dimensions other than four. We obtain reliable
values for the scaling exponent in the supercritical region for dimensions in the
range $3.5\leq D\leq 14$. The critical exponent increases monotonically to an asymptotic value at large $D$ of
$\gamma\sim0.466$. The data is well fit by a simple exponential of the form:
$\gamma \sim 0.466(1-e^{-0.408 D})$.

\end{abstract}


\maketitle

\section{Introduction}
Choptuik\cite{choptuik} was first to discover that gravitational
collapse to black holes exhibits critical behaviour\cite{gund,lehn}.
In particular, if one evolves initial data corresponding to a matter
distribution described by some parameter $a$, there is a critical
value $a_*$ of that parameter that separates data whose final state
is a black hole, from that for which the matter ultimately disperses
to infinity. Near the critical value, the radius of
the horizon on formation obeys the simple scaling
law\cite{choptuik}: \beq R_{ah}(a)=(a-a_*)^\gamma \eeq The scaling
exponent $\gamma$ depends only on the type of matter that is
considered (eg scalar field, vector field, etc.) and not on the
choice of parameter $a$ or the distribution of the initial data. The
emergence of this universal exponent is fairly well understood
numerically in terms of a unique, discretely self-similar critical
solution which serves as an attractor. The self-similarity is
manifested by a periodicity of the critical solution for the scalar
field $\chi(r,t)$ of the form: \beq \chi(r,t)=\chi(r e^\Delta,
te^\Delta) \eeq with echoing period $\Delta$. The critical solution
has only been studied numerically and very little is known analytically
about it, except perhaps in the context of the three
dimensional BTZ black hole\cite{DG3}.

Until recently, little was known about critical gravitational
collapse in spacetime dimensions other than four. The only
exceptions were in three\cite{pc, husain} and six
dimensions\cite{DG_6d}. A few years ago, two of the current authors
started a program \cite{d_dim1} designed to examine the dependence of the critical exponent on spacetime dimension.  A formalism was set up that allowed
the use of a single code in
which dimension appeared as an input parameter. Previous
results in four and six dimensions were confirmed and some
preliminary new results in five dimensions were presented. More
recently, motivated in part by the relationship between critical
phenomena in gravitational collapse and black hole to black string
phase transitions in higher dimensions, Sorkin and Oren\cite{sorkin}
performed a similar analysis using a different formalism.

In the following, we describe 
the dependence of the critical exponent $\gamma$ on
spacetime dimension. A simple, but significant, improvement
to the original code
in \cite{d_dim1} has enabled us to obtain reliable and accurate
results for dimensions $3.5\leq D\leq14$ \footnote{Since spacetime
dimensions appears algebraically as a parameter in the code,
mathematically there is no impediment to considering non-integer
dimension, despite the fact that it has no obvious physical
interpretation. The equations are however singular at $D=3$.}.  In this range
the critical exponent grows monotonically with $D$ to an asymptotic
value near $0.466$. Remarkably, the data is well fit by a simple exponential:
\beq
\gamma = 0.466(1-e^{-0.408 D})
\label{gamma fit 1}
\eeq

Our data are consistent with those of \cite{sorkin}. Our error bars are however smaller, and we are able to get reliable results in higher dimensions. We see no evidence for the maximum
in $\gamma$ near $D=11$ suggested in \cite{sorkin}.

We first review our general formalism, then describe the numerical
method and calculation of error. Finally, we present our results and
conclusions.

\section{Formalism}

We begin with Einstein gravity in $D$ spacetime dimensions minimally
coupled to a massless scalar field: \bea S^{(D)}&=&{1\over 16\pi
G^{(D)}}\int d^Dx\sqrt{-g^{(D)}}
   \left[R(g^{(D)}) - \Lambda\right] \nonumber \\
   && - \int d^Dx\ \sqrt{-g^{(D)}}|\partial\chi|^2.
\label{Einstein}
\eea

Spherical symmetry is imposed by writing the metric $g_{\mu\nu}$ as
\be
ds^2_{(D)} = \bar{g}_{\alpha\beta} dx^\alpha dx^{\beta} + r^2(x^\alpha)
d\Omega_{(D-2)},
\label{metric 1}
\ee
where $d\Omega_{(D-2)}$ is the metric on $S^{D-2}$ and $\alpha,\beta =
1,2$.

The key to obtaining a simple and elegant set of equations in all
dimensions is a field redefinition motivated by 2-d dilaton
gravity\cite{dil_grav,d_dim1}: \bea
\phi &:=& {n\over 8(n-1)}\left({r\over l}\right)^n, \\
g_{\alpha\beta} &:=& \phi^{(n-1)/n}\ \bar{g}_{\alpha\beta},
\label{defs} \eea where $n\equiv D-2$. Note that $r$ is the optical
scalar, and $\phi$ is proportional to the area of the $n$-sphere at
fixed radius $r$. Formation of an apparent horizon in such a
spherically symmetric spacetime is signalled by the vanishing of the
quantity: \be ah\equiv {g}^{\alpha\beta}\partial_\alpha
 \phi \partial_\beta \phi,
\label{aheqn}
\ee

We go to double null coordinates first introduced by Garfinkle\cite{DG}:
\beq
ds^2 = - 2 l g(u,v)\phi'(u,v) du dv
\label{double null}
\eeq
in which the relevant field equations take the simple form:
\bea
& &\dot{\phi}' = - {l\over 2} V^{(n)}(\phi) g \phi'
\label{double null equations a}\\
& &{g'\phi'\over g \phi} = 32\pi (\chi')^2 \label{cons}\\
& &(\phi\chi')^{\cdot} + (\phi\dot{\chi})' = 0,
\label{double null equations c}
\eea
where prime and dot denote the $v$ and $u$ derivatives respectively.
All the information about spacetime dimension is encoded in the
potential $V^{(n)}$:
\bea
V^{(n)}(\phi,\Lambda) &\equiv& (n-1)\left[
\tphi^{-1\over n}
 -l^2 \Lambda \tphi^{1/n}\right]
\label{poten}
\eea
with:
\beq
\tphi\equiv {8(n-1)\over n} \phi = \left({r\over l}\right)^n
\eeq
 As one might expect from the fact that critical collapse
 is a short distance phenomena, the cosmological constant
 does not affect the critical exponent. This was verified
 by explicit calculation in \cite{d_dim2}. We will henceforth
 set $\Lambda$ to zero and concentrate on the dependence of $\gamma$ on
 spacetime dimension $D$.

The evolution equations may be put in a more useful form
 by defining the variable
\be h = \chi + {2\phi\chi' \over \phi'},
\label{def h}
\ee
which replaces the scalar field $\chi$ by $h$. The evolution equations become
\bea
\dot{\phi} &=& -\tilde{g}/2 \label{phidot}\\
\dot{h} &=& {1\over 2\phi} (h - \chi)\left( g\phi V^{(n)} - {1\over 2}\tilde{g}\right),
\label{hdot}
\eea
\bea
g &=& \exp\left[4\pi\int^v_udv{\phi'\over\phi}(h-\chi)^2\right],
   \label{g}  \\
\tilde{g} &=& l\int _u^v (g \phi'V^{(n)}) dv,
\label{gbar}
\eea
\be
\chi={1\over
2\sqrt{\phi}}\int^v_udv\left[{h\phi'\over\sqrt{\phi}}\right].
\label{chi integral}
\ee

This was the final form of the equations used for numerical
evolution in \cite{d_dim1}. Note, however, that the integrand of
$\tilde{g}$ is singular in the critical region near $\phi=0$, even
though the integral itself is well behaved. This leads to
instabilities that severely limit the accuracy of the calculation. A
drastic improvement can be obtained by observing that: \bea
\int^v_u g V^{(n)} \phi' dv &=& (n-1)\left({8n\over n-1}\right)^{-1/n}
 \int^v_u g \phi^{-1/n}d\phi \nonumber\\
&=& n\left({8n\over n-1}\right)^{-1/n} \int^v_u g d\phi^{1-1/n}
\eea
 so that a straightforward integration by parts yields:
\be
\tilde{g}={n\over n-1}g \phi V^{(n)}-{4\pi n\over n-1}\int^v_u g V^{(n)}
 (h-\chi)^2 \phi' dv
\label{new gtilde} \ee 
where we have used the expression for $g$ in
(\ref{g}) to write: 
\be dg = 4\pi g (h-\chi)^2 {\phi'\over \phi} dv
\ee 
The first term in (\ref{new gtilde}) vanishes at $\phi=0$, and
the integrand in the second term now vanishes at the origin by virtue of
(\ref{def h}) (assuming $\phi' \neq 0$). Thus the use of (\ref{new gtilde}) instead of (\ref{gbar}) improves the stability of the code near criticality significantly and allows reliable results to be obtained in dimensions up to at least 14.

\section{Numerical Method}
The numerical scheme uses a $v$ (`space') discretization to obtain a
set of coupled ODEs: \be h(u,v) \rightarrow h_i(u),\ \ \ \ \ \ \
\phi(u,v)\rightarrow \phi_i(u). \ee where $i = 0,\cdots, N$
specifies the $v$ grid. Initial data for these two functions is
prescribed on a constant $v$ slice, from which the functions
$g(u,v),\tilde{g}(u,v)$ are constructed. Evolution in the `time'
variable $u$ is performed using the $4^{th}$ order Runge-Kutta
method. The general scheme is similar to that used in \cite{GP},
together with some refinements used in \cite{DG}. This procedure was
also used for the $3-$dimensional collapse calculations in
\cite{husain}. The grid size used for all calculations reported here
was 6000, with no mesh refinement. An improvement in  accuracy was, however, obtained by adaptively decreasing the time steps once apparent horizon
formation commenced, i.e. once the function $ah(\phi)$ developed an
extremum. Time steps ranged from $\Delta u = 10^{-3}$ at the
beginning of the calculation to about $\Delta u = 10^{-9}$ near the
end.

The boundary conditions at fixed $u$ are \be \phi_k=0, \ \ \
\tilde{g}_k=0,\ \ \ \ g_k=1. \ee {where $k$ is the index
corresponding to the position of the origin $\phi=0$.} All grid
points $ 0\le i \le k-1$ correspond to ingoing rays that have
reached the origin and are dropped from the grid. The boundary
conditions are equivalent to $r(u,u)=0$, $g|_{r=0} = g(u,u) =1$.
Notice that for our initial data, $\phi_k$ and hence $h_k$, are
initially zero, and therefore remain zero at the origin because of
(\ref{hdot}). The initial scalar field configuration
$\chi(u=0,\phi)$ is most conveniently specified as a function of
$\phi$ rather than $r$. (Recall that $\phi\propto r^n$.)  This
together with the initial arrangement of the radial points
$\phi(0,v)$ fixes all other functions. We used the initial
specification $\phi(0,v) = v$.

We considered Gaussian initial data \be \chi_G(u=0,\phi) = a
\phi^{2/n}\ {\rm exp}\left[-\left(\ {\phi-\phi_0\over
\sigma}\right)^2\right], \ee The initial values of the other
functions were determined in terms of the above by computing the
integrals for $g_n$ and $\tilde{g}_n$ using Euler's method for
equally spaced points.

At each $u$ step, a check is made to see if an apparent horizon has
formed by observing the function $ah$ in (\ref{aheqn}) whose
vanishing signals the formation of an apparent horizon. For each run
of the code with fixed parameters $a$ and $\sigma$, this function is
scanned from larger to smaller radial values after each  Runge-Kutta
iteration for the presence of an apparent horizon.  In the
subcritical case, it is expected that all the radial grid points
reach zero without detection of an apparent horizon. This is the
signal of pulse reflection. In the supercritical case, black hole
formation is signalled by $ah$ vanishing at some finite radius.
Since the code crashes before $ah=0$ is actually reached, one has to
terminate it when $ah(\phi_{min})$ is below some bound. Note that by virtue of our parameterization, in vacuum $ah$ decreases monotonically to zero
at $r=0$. As the radius of horizon formation decreases, it is
therefore necessary to decrease the value of the bound. The required
values ranged from about $10^{-6}$ near criticality in lower
dimensions to $10^{-34}$ in higher dimensions.

\begin{figure}[ht]
\begin{center}
\epsfxsize=80mm \epsffile{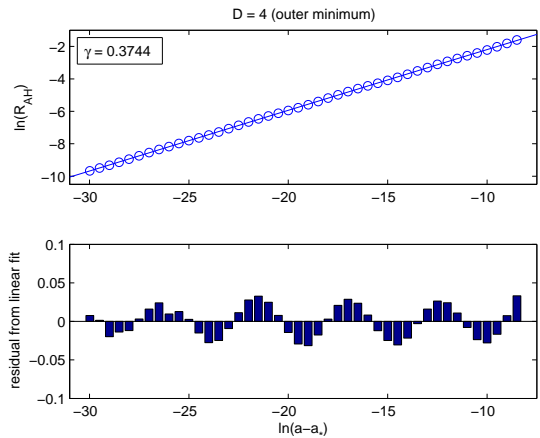} \caption{Plot of
$ln(R_{AH})$ vs. $ln(a-a_*)$ for $4$ dimensional case. The residuals
clearly show the oscillation expected due to discrete
self-similarity of the critical solution.} \label{4d_figure}
\end{center}
\end{figure}

In all cases, we used values of $\phi_0=1$ and $\sigma =0.3$ for the
Gaussian initial data and varied the amplitude $a$ to determine the
critical amplitude $a_*$ below which no black hole formation was
observed. We then did runs in the supercritical region $a>a_*$ to
determine the dependence of the horizon radius $R_{ah}$ on the
amplitude. The critical exponent is given by the slope of the best
fit straight line for the $\ln(R_{ah})$ vs $\ln(a-a_*)$ plot, while
the echoing period is determined from the periodicity of the
residuals. See Fig.(\ref{4d_figure}).

\begin{figure}[ht]
\begin{center}
\epsfxsize=80mm \epsffile{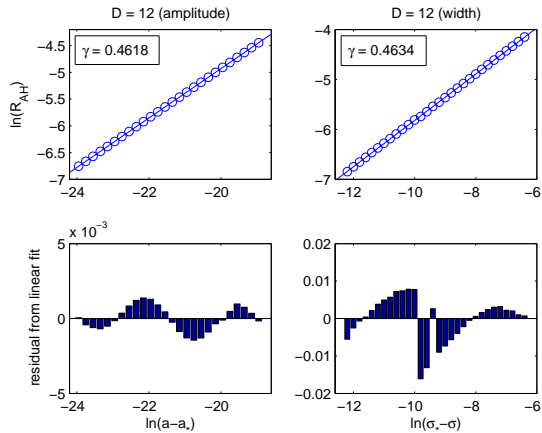} \caption{Plot of
$ln(R_{AH})$ vs. $ln(a-a_*)$ and $ln(R_{ah})$ vs.
$ln(\sigma_*-\sigma)$ for our $12$ dimensional case. Although the
residuals are cleaner in the amplitude case, the slopes of the lines
agree to within error.} \label{12d_figure}
\end{center}
\end{figure}

Universality was tested by fixing $a=a_*$ in the program, treating
$\sigma=\sigma_*=0.3$ as the critical width, and then varying the
width in the supercritical region $\sigma<\sigma_*$. In all cases
that we tested, varying the width and varying the amplitude gave the
same critical exponents to the required accuracy. For example in 12
dimensions (see Fig.(\ref{12d_figure})) the difference between the two
values was less than 0.5\%.

The accuracy of the critical exponent is in principle
limited by the number of significant digits in the critical
amplitude. Due to instabilities in the code near criticality, it
became progressively more difficult to determine $a_*$ directly as
the dimension increased. Our error estimates were determined as follows. We first determined a range for $a_*$ by
direct search. We then varied $a_*$ in the determined range, and for each value we calculated and plotted $ln(R_{AH})$ vs
$ln(a-a_*)$. The value of $a_*$ used to determine $\gamma$ as quoted in Table (1) was the one for which the average squared deviation $R^2$ of the $ln(R_{AH})$ vs
$ln(a-a_*)$ plot was minimized.  Since the tail of the $ln(R_{AH})$ vs
$ln(a-a_*)$ plot is fairly sensitive to the
value of $a_*$, we were able to narrow the range of $a_*$, and hence
lower the error bar on the slope, by determining the values of $a_*$
for which $R^2$ was within a specified range of its maximum (generally $\pm 0.0001$). Although there was some variation with dimension, the procedure generally gave us an error in $\gamma$ of about 1\%. The error was less in 4 dimensions, for which the code was most stable. It is interesting that the overall error did not grow rapidly at high dimension, despite the fact that our determination of $a_*$ became less accurate.

It is also important to note, that in all
cases tested, including $14$ dimensions, varying the width
of the initial matter distribution yielded the same value for the
critical exponent as varying the amplitude, within the given error
bars. See for example Fig.(\ref{12d_figure}).

\begin{figure}[ht]
\begin{center}
\epsfxsize=80mm \epsffile{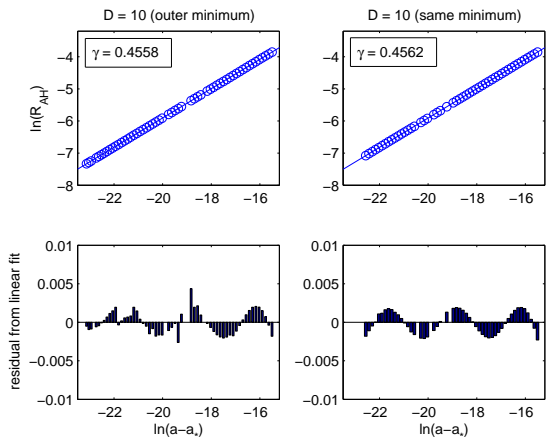} \caption{Plot of
$ln(R_{AH})$ vs. $ln(a-a_*)$ for $D=10$. The curve on the left
tracks the radius of the true horizon on formation, whereas the
curve on the right tracks the same minimum throughout. The slopes
are within error, but the residuals on the right clearly give the
echoing period associated with discrete self-similarity.}
\label{10d_figure}
\end{center}
\end{figure}

\begin{figure}[ht]
\begin{center}
\epsfxsize=80mm
\epsffile{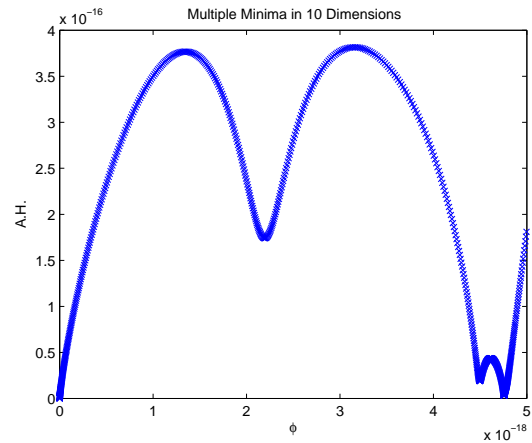}
\caption{Plot of ah vs. phi for 10 dimensions, showing the appearance of multiple minima. The outermost minimum corresponds to the horizon location.}
\label{mult_min}
\end{center}
\end{figure}

The echoing period can be obtained in a straightforward way by measuring the
period of oscillations of the residuals from the straight line in
the ln-ln plots. It is given by: \beq \Delta = 2 T
\gamma \eeq where $T$ is the oscillation period. The accuracy is
limited by the errors in both $\gamma$ and $T$ , and is therefore
considerably worse than that of the critical exponent. The
determination of $\Delta$ is further complicated in higher
dimensions by the appearance of multiple minima (see
Fig.(\ref{mult_min})). 

Although the horizon location is always given by the
outermost minimum the recorded value jumps as new minima
appear. When the location of a single minimum (not necessarily the
outer one which formed the horizon) is tracked throughout the range
of amplitudes considered, the residuals are much smoother, and give
a more accurate indication of the echoing period. This procedure does
not significantly affect the slope (see Fig.(\ref{10d_figure})).

\begin{figure}[ht]
\begin{center}
\epsfxsize=80mm
\epsffile{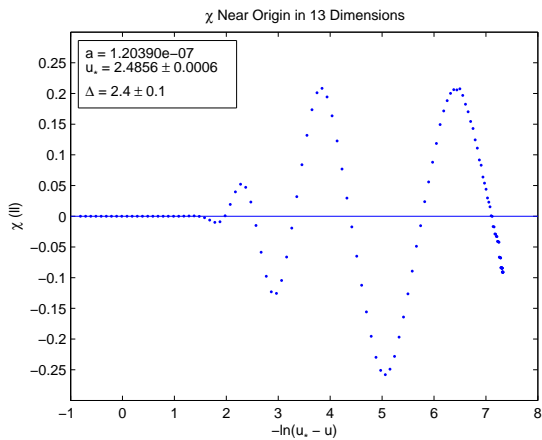}
\caption{Plot of $\chi$ vs. $\ln(u-u_*)$ near criticality for 13 dimensions, showing the ringing expected from the discrete self-similarity of the solution.}
\label{ringing}
\end{center}
\end{figure}

Another, potentially more accurate, estimate for $\Delta$ can be obtained by observing the ringing of the scalar field near the origin just prior to the formation of an apparent horizon near criticality\footnote{We are grateful to E. Sorkin for helpful conversations in this regard.}. Although our graphs gave a clear indication of the ringing (see Fig. (\ref{ringing}), we were unable to get close enough to criticality to obtain very accurate results for the period. This quantity is generally easier to obtain by observing the scalar curvature in the sub-critical region, as done in \cite{sorkin}. We estimate that our error in determining $\Delta$ was in the neighbourhood of $\pm 0.1$.

\section{Results and Discussion}

The results for the critical exponent and echoing period are given
in Table (1) below. The universality of the results, plus the clear
presence of oscillations in the ln-ln plots confirm that we are
sufficiently close to criticality in all cases to obtain reliable
results.

\begin{table}[h]
\begin{tabular}{||c||c|c||c|c||}
\hline
D& $\gamma$&$\gamma$ (Ref.\cite{sorkin}) & $\Delta (\pm 0.1)$ &$\Delta$ (Ref.\cite{sorkin})\\
 \hline 
 \hline 3.5& $0.349\pm 0.003$ &  & & \\ 
\hline 4& $0.374\pm 0.002$ &$0.372 \pm 0.004$ & $3.40$ &$3.37 \pm 2\%$ \\
\hline 5& $0.412\pm 0.004$ & $0.408 \pm 0.008$&$3.10$&$3.19 \pm 2\% $ \\
\hline 6& $0.430\pm 0.003$ & $0.422 \pm 0.008$&$2.98$&$3.01 \pm 2\%$ \\
 \hline 7& $0.441\pm 0.004$ & $0.429 \pm 0.009$&$2.96$&$2.83 \pm 2\%$ \\ 
\hline 8& $0.446\pm 0.004$ & $0.436 \pm 0.009$&$2.77$&$2.70 \pm 3\%$ \\ 
\hline 9& $0.453\pm 0.003$ & $0.442 \pm 0.009$&$2.63$&$2.61 \pm 3\%$  \\ 
\hline 10& $0.456\pm0.004$ & $0.447 \pm 0.013$&$2.50$&$2.55 \pm 3\%$ \\
\hline 11& $0.459\pm 0.004$ & $0.44 \pm 0.013$&$2.46$&$2.51 \pm 3\%$ \\ 
\hline 12& $0.462 \pm 0.005$ &  &2.44 & \\ 
\hline 13& $0.463 \pm 0.004$ &  &2.40 & \\
\hline 14& $0.465 \pm 0.004$ &  &  & \\ \hline
 \hline
\end{tabular}
\caption{Results for critical exponent and echoing period.
Universality of the critical exponent was checked in several
dimensions by varying width and amplitude. Comparisons with previous work
of \cite{sorkin} is given in columns 3 and 5. }
\end{table}

\begin{figure}[ht]
\begin{center}
\epsfxsize=80mm \epsffile{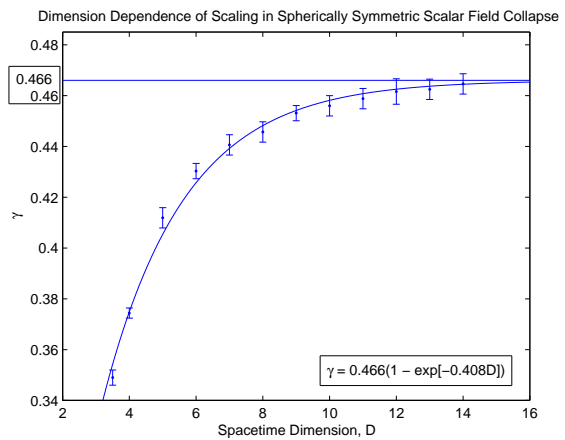} \caption{Plot of results for
$\gamma$ as a function of spacetime dimension.} \label{gamma_fig}
\end{center}
\end{figure}

The critical exponent appears to grow with dimension monotonically
to an asymptotic value near 0.466.  As seen in
Fig.(\ref{gamma_fig}), the data are well fit by an exponential of
the form: $
\gamma = A-B e^{-C\times D}
\label{exponential fit}
$
with $A\sim B\sim 0.466$ and $C\sim 0.408$. The fact that $A$ and $B$ are approximately equal is significant, since they are {\it a priori} independent parameters.

Our results are consistent within error with those obtained by Sorkin and Oren\cite{sorkin}, for both the critical exponent $\gamma$, and the self-similar period $\Delta$. However, we do not observe the maximum for the critical exponent near $D=11$ reported in \cite{sorkin}. There is no contradiction here, since the authors of that paper calculated the critical exponent to $11$ dimensions, and conjectured a maximum by extrapolating polynomial fits to higher dimensions. In this context it may also be useful to note that a maximum in the critical exponent was also present in the preliminary results reported by one of the authors in \cite{GK_dublin}. In our case, the maximum was determined to be an artefact of the inaccuracy of the calculation. At higher dimensions it is more difficult to obtain numerical stability near criticality. Moreover, the slope of the plot of $ln(R_{ah})$ vs. $\ln(a-a*)$ systematically decreases as you move further away from $a*$. The critical exponents reported in \cite{GK_dublin} 
decreased for dimensions greater than ten because the data were generated by points progressively further away from criticality.

Our results for the ringing period $\Delta$ suggest that $\Delta$ decreases monotonically to an asymptotic value of just below 2.4.  This is not unreasonable given our more reliable results for the critical exponent. If indeed the $D=\infty$ limit of the questions is well defined, then it is reasonable that both $\gamma$ and $\Delta$ asymptote to a finite value determined by the critical solution of the limiting theory. It is of great interest to try to find this solution analytically. In fact one of the original goals that motivated the program started in \cite{d_dim1} was the hope that a numerical analysis of the dimension dependence of critical collapse would provide clues about the nature of this limiting solution. This is currently under further investigation.

\begin{figure}[ht]
\begin{center}
\epsfxsize=80mm \epsffile{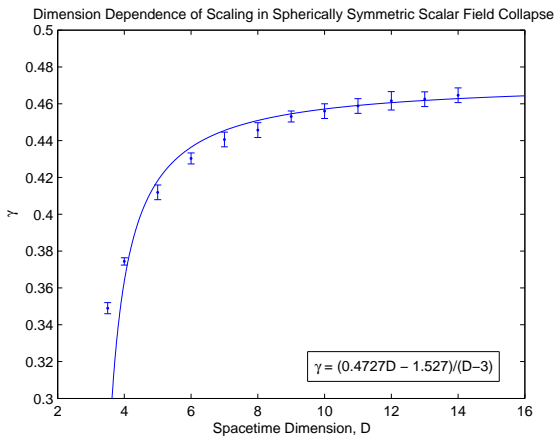} \caption{Alternative fit for
$\gamma$ as a function of spacetime dimension that diverges at $D=3$.} \label{gamma fit 2}
\end{center}
\end{figure}

In summary, we have presented the results of an analysis of the dependence of the critical exponent in spherically symmetric scalar field collapse on spacetime dimension. There is strong evidence that this exponent increases monotonically to an asymptote around $0.466$. The data are remarkably well fit by the simple relationship (\ref{gamma fit 1}), which also suggests that the critical exponent is finite at $D=3$. However, one can also fit the data, albeit not quite as well, (see Fig.(\ref{gamma fit 2})), with an expression of the form: 
\beq \gamma = 0.4727 - \left({1.527\over D-3}\right) 
\eeq 
which goes to minus infinity at $D=3$. In order to settle this point definitively we need to explore the region near $D=3$ more thoroughly. This work is currently in progress..

Given the recent interest in higher dimensions in general, and
higher dimensional black holes in particular it is worthwhile to
understand in detail the nature of gravitational collapse to black
holes in all dimensions. This may be particularly relevant in the
context of black string to black hole transitions in higher
dimensions. The results presented here should provide valuable clues
in addressing these issues.

\section{Acknowledgements}
This work was supported in part by the Natural Sciences and Engineering
Research Council of Canada. The authors are grateful to David Garfinkle and Evgeny Sorkin for invaluable discussions and to the Perimeter Institute for hospitality.



\begin{thebibliography}{99}

\bibitem{choptuik} M. Choptuik, Phys. Rev. Lett. {\bf 70}, 9 (1993).

\bibitem{gund} C. Gundlach, Living Rev. Rel. 2, 4 (1999).

\bibitem{lehn} L. Lehner, Class. Quant. Grav. 18,  R25-R86 (2001)

\bibitem{DG3} D. Garfinkle, Phys.Rev. {\bf D63} (2001) 044007.

\bibitem{pc} F. Pretorius and M. Choptuik, Phys. Rev. {\bf D62} (2000) 124012.

\bibitem{husain} V. Husain and M. Olivier, Class. Quant. Grav {\bf 18}
L1-L10 (2001).

\bibitem{DG_6d} D. Garfinkle, C. Cutler, G. Comer Duncan Phys.Rev. D60 (1999) 104007.

\bibitem{d_dim1} M. Birukou, B. Husain, G. Kunstatter, E. Vaz and M. Olivier,
Phys. Rev. {\bf D65}, 104036 (2002).

\bibitem{sorkin}  E. Sorkin, Y. Oren, Phys.Rev. D71 (2005) 124005.

\bibitem{dil_grav}J. Gegenberg, D. Louis-Martinez and G. Kunstatter,
Phys. Rev. {\bf D51} 1781 (1995); D. Louis-Martinez and G. Kunstatter,
Phys. Rev. {\bf D52} 3494 (1995).

\bibitem{DG2} D. Garfinkle and  G. Comer Duncan, Phys. Rev. {\bf D58} (1998) 064024.



\bibitem{DG} D. Garfinkle, Phys. Rev. {\bf D51}, 5558 (1995).

\bibitem{d_dim2} V. Husain, G. Kunstatter, B. Preston, M. Birukou, Class.Quant.Grav. 20 (2003) L23-L30.

\bibitem{GP} D. Goldwirth and T. Piran, Phys. Rev. {\bf D36}, 3575
(1987).



\bibitem{GK_dublin} G. Kunstatter,  ``Dimension Dependence of the Critical Exponent in Spherical Black Hole Formation'', contributed talk given at GR17, Dublin, July, 2004.



%
\end{thebibliography}
\end{document}